\documentclass[twocolumn,twocolumnappendix]{aastex631}

\usepackage{xcolor}
\usepackage{booktabs}
\usepackage{upgreek}
\usepackage{comment}

\begin{document}

\title{Constraints on the Orbit of the Young Substellar Companion GQ Lup B from High-Resolution Spectroscopy and VLTI/GRAVITY Astrometry \footnote{Based on observations collected at the European Southern Observatory under ESO programmes 1104.C-0651 and 109.238N.001.}}

\author[0000-0002-5638-4344]{Vidya~Venkatesan}
\altaffiliation{NASA FINESST Fellow}
\affiliation{Department of Physics and Astronomy, University of California, Irvine, California, USA}

\author[0000-0002-3199-2888]{S.~Blunt}
\affiliation{Department of Physics and Astronomy, University of California, Santa Cruz, California, USA}

\author{J.~J.~Wang}
\affiliation{Center for Interdisciplinary Exploration and Research in Astrophysics (CIERA) and Department of Physics and Astronomy, Northwestern University, Evanston, IL 60208, USA}
\author{S.~Lacour}
\affiliation{LESIA, Observatoire de Paris, PSL, CNRS, Sorbonne Universit\'e, Universit\'e de Paris, 5 place Janssen, 92195 Meudon, France}
\affiliation{European Southern Observatory, Karl-Schwarzschild-Stra\ss{}e 2, 85748 Garching, Germany}
\author[0000-0002-2919-7500]{G.-D.~Marleau}
\affiliation{Division of Space Research \&\ Planetary Sciences, Physics Institute, University of Bern, Sidlerstr.~5, 3012 Bern, Switzerland}
\affiliation{Max-Planck-Institut f\"ur Astronomie, K\"onigstuhl 17, 69117 Heidelberg, Germany}
\affiliation{Fakult\"at f\"ur Physik, Universit\"at Duisburg-Essen, Lotharstra\ss{}e 1, 47057 Duisburg, Germany}
\author{G.A.L.~Coleman}
\affiliation{Astronomy Unit, Department of Physics and Astronomy, Queen Mary University of London, Mile End Road, London, E1 4NS, UK}
\author{L.~Guerrero}
\affiliation{Department of Physics and Astronomy,
Northwestern University, Evanston, IL 60208, USA}

\author{W.~O.~Balmer}
\affiliation{Department of Physics \& Astronomy, Johns Hopkins University, 3400 N. Charles Street, Baltimore, MD 21218, USA}
\affiliation{Space Telescope Science Institute, 3700 San Martin Drive, Baltimore, MD 21218, USA}
\author{L.~Pueyo}
\affiliation{Space Telescope Science Institute, 3700 San Martin Drive, Baltimore, MD 21218, USA}
\author{T.~Stolker}
\affiliation{Leiden Observatory, Leiden University, P.O.\ Box 9513, 2300 RA Leiden, The Netherlands}
% ...

\begin{abstract}
Understanding the orbits of giant planets is critical for testing planet formation models, particularly at wide separations ($>$10~au) where traditional core accretion becomes inefficient. However, constraining orbits at these separations has historically been challenging due to sparse orbital coverage and related degeneracies in the orbital parameters. In this work, we use existing high-resolution ($R \sim 100{,}000$) spectroscopic measurements from CRIRES+, astrometric data from SPHERE, NACO, and ALMA, and combine it with new high-precision GRAVITY astrometry data to refine the orbit of GQ~Lup~B, a $\sim$30~$M_{\mathrm{J}}$ companion at $\sim$100~au, in a system that also hosts a circumstellar disk and a wide companion, GQ~Lup~C. Including RV data significantly improves orbital constraints by breaking the degeneracy between inclination and eccentricity that plagues astrometry-only fits for long-period companions. Our work is one of the first to combine high-precision astrometry with the companion’s relative radial velocity measurements to achieve significantly improved orbital constraint. The eccentricity is refined from $e = 0.47^{+0.14}_{-0.16}$ (GRAVITY only) to $e = 0.35^{+0.10}_{-0.09}$ when RVs and GRAVITY data are combined. We also compute the mutual inclinations between the orbit of GQ~Lup~B, the circumstellar disk, the stellar spin axis, and the disk of GQ~Lup~C. The orbit is misaligned by $63^{+6}_{-14}$$^\circ$ relative to the circumstellar disk, $52^{+19}_{-24}$$^\circ$ with the host star's spin axis, but appears more consistent ($34^{+6}_{-13}$$^\circ$) with the inclination of the wide tertiary companion GQ~Lup~C's disk. These results support a formation scenario for GQ~Lup~B consistent with cloud fragmentation. They highlight the power of combining companion RV constraints with interferometric astrometry to probe the dynamics and formation of wide-orbit substellar companions.
\end{abstract}


\section{Introduction} \label{sec:intro}

One of the the fundamental puzzles in exoplanetary science is the formation and evolution of giant planets and substellar companions at wide separations (tens to hundreds of~au away from their host stars). At such distances, the traditional core accretion theory struggles to explain the formation of a planetary core massive enough to trigger runaway gas accretion before the dissipation of the protoplanetary disk \citep{Pollack1996,Inaba2003,Ikoma2025}. Gravitational instability in the disk is one alternative formation pathway that consists of massive clump formation at $>$ 100~au \citep{Boss1997, Mayer2002}, followed by inward migration and potential disruption \citep{Kratter2016,Nayakshin2017}. Alternative scenarios such as dynamical scattering \citep{Veras2009}, outward migration via disk--planet interactions \citep{Crida2009}, and direct collapse out of molecular cloud \citep{Stamatellos2007, Stam&Whitworth2009} have also been proposed. Each pathway is expected to leave a unique imprint on a system's dynamical architecture, particularly in orbital eccentricities \citep{Marleau2019}, mutual inclinations (difference in orbital angular momentum vector orientations of planets in the same system; \citealp{Rosa2020}), and spin--orbit obliquities (difference in vector orientations of stellar spin axes and planetary orbit; \citealp{Bowler2023, Bryan2021}). Therefore, constraining the orbits of exoplanets with limited astrometric temporal baselines may be key to uncovering the formation mechanisms of widely separated objects. However, for most wide-orbit companions, these dynamical signatures remain poorly constrained due to long orbital periods, typically 1000+ years, and corresponding limited observational baselines. 

The GQ~Lup system is located in the Lupus star-forming region approximately 151 parsecs away \citep{GAIADR32023}, and consists of a young (\textless 2-5~Myr) \citep{Schwarz2016, Donati2012, Stolker2021}, K7 T Tauri star, a transition disk extending to $\sim$50~au \citep{MacGregor2017, Wu2017}, and a directly imaged substellar companion, GQ Lup B. One of the first directly-imaged substellar companions \citep{Neuh2005}, with a projected separation of $\sim$100~au and an orbital period of $\sim$900-1200 years \citep{Stolker2021}, GQ~Lup~B has an estimated mass between 10--40~$M_{\mathrm{J}}$ \citep{Stolker2021}, placing it near the planet/brown dwarf boundary, (which may occur around 15--40~$M_{\mathrm{J}}$ based on features in the stellar mass function \citep{Chabrier2003,Reggiani2016,Stevenson2023}). In addition to GQ~Lup~B, the system hosts a likely wide tertiary companion, GQ~Lup~C, identified by \citet{Lazzoni2020, Alcala2020} at a projected separation of $\sim$2400~au. Its bound status remains uncertain, but such a distant companion could play a role in the long-term dynamical evolution of the system. With two companions at wide separations and the presence of a circumstellar disk, this system offers a valuable laboratory to investigate the formation and evolution of widely separated objects. 

Current observational evidence from GQ~Lup~B’s atmosphere supports a formation scenario via disk instability or direct collapse. High-resolution spectroscopic studies have revealed CO and H$_2$O absorption features in its atmosphere \citep{Schwarz2016}, and its carbon isotopic ratio and C/O ratio are consistent with those of the host star \citep{Gonzalez2025,Xuan2024}, favoring a common origin and suggesting formation from the same protostellar disk or via gravoturbulent fragmentation of a single molecular cloud. Moreover, detection of strong H$\alpha$ emission indicates that the planet is still accreting \citep{Marois2007}, and the possible presence of a circumplanetary disk \citep{Cugno2024} further supports the interpretation of GQ~Lup~B as a young, evolving system with the potential for satellite formation and continued dynamical evolution.

In addition to the atmospheric composition, orbital architecture offers a complementary avenue for probing the origins of substellar companions (e.g., \citealt{Xuan2022}). Precise orbital parameters—particularly eccentricity—can help differentiate between formation scenarios, since high eccentricity can indicate  scattering \citep{Dong2016, Bryan2016}, while low eccentricity can imply disk driven migration, or in-situ formation consistent with the overall eccentricity distribution of directly-imaged planets \citep{ Bowler2020, Nagpal2023, Do2023}. Historically, the orbital characterization of directly imaged planets has relied almost exclusively on relative astrometry, limiting our ability to tightly constrain orbital elements—particularly eccentricity and inclination. However, widely separated systems like GQ~Lup~B pose a challenge for traditional orbit-fitting techniques due to their long orbital periods. For this companion, only about 0.5\% of the full orbit is currently sampled by the astrometric baseline, limiting the precision with which its orbital parameters can be constrained. This limitation is amplified for long-period companions like GQ~Lup~B, where current astrometric baselines cover only a small fraction of the orbit, and measurements from absolute astrometry with Hipparcos and/or Gaia are unconstraining. A key challenge in such cases is the well-known degeneracy between eccentricity and inclination \citep{Ferrer-Chavez2021}, where nearly circular, edge-on orbits can be difficult to distinguish from more eccentric, more face-on configurations based solely on sky-projected motion. Incorporating high-precision relative RV data, even over a relatively short baseline, can help break this degeneracy. Rather than waiting decades to detect orbital curvature via astrometry alone, precise RV measurements can improve constraints \citep{Blunt2023}.

The advent of high-precision astrometry from instruments like GRAVITY has enabled robust orbital constraints for directly imaged companions, including HIP99770~b \citep{Winterhalder2025}, HD136164~Ab—an eccentric brown dwarf likely formed via fragmentation \citep{Balmer2024}—AF~Lep~b, whose circular orbit is consistent with a core accretion scenario \citep{Balmer2025}, and YSES1~b (Roberts et al., in prep.). Earlier efforts also demonstrated GRAVITY’s power in systems such as HD206893~b \citep{Kammerer2021} and $\beta$~Pic~b \citep{GRAVITY2020,Brandt2021bPic}. This progress is further complemented by advances in high-resolution spectroscopy. Instruments like CRIRES+ can now allow us to place a slit directly on GQ~Lup~B and extract planetary radial velocity. This technique, long a cornerstone of stellar orbital characterization, is now viable for planets on wide-separation from their host stars and with favorable contrasts. 

While \citet{Stolker2021} combined RV data with VLT/NACO and VLT/MUSE astrometry, the low-precision measurements resulted in large uncertainties. This study combines multi-epoch astrometric measurements, including new high-resolution astrometry from the GRAVITY instrument, with recent planetary RV measurements derived from the spectrum of GQ~Lup~B itself to refine the orbital parameters of the companion and assess their implications for its formation. In addition to refining the orbital parameters of GQ Lup B, this work evaluates the relative information content of the available observational datasets, focusing on relative astrometry and relative RVs, in constraining GQ Lup B's orbit. These orbit fitting methods are also directly relevant to the science goals of future missions such as the \textit{Habitable Worlds Observatory} (HWO; \citealp{HWO2023}). In particular, this technique of combining high-precision astrometry with relative RVs to break the degeneracies between inclination and eccentricity could potentially be adapted to HWO rocky planets. In addition to gaining insights into the formation and dynamical history of the system, accurate orbital solutions are also critical for determining key parameters (e.g., orbital phase and stellar irradiation) that inform climate modeling and habitability assessments for eccentric rocky planets with HWO~\citep{Venkatesan2025}.

The paper is structured as follows. Section~\ref{sec:data} describes the data acquisition and inputs used in \texttt{orbitize!}. Section~\ref{sec:orbit} outlines our orbit-fitting methodology and the calculation of orbital eccentricity using both radial velocity (RV) and astrometric data. The results are presented in Section~\ref{sec:res}, and discussed in Section~\ref{sec:dis}. Finally, we summarize our conclusions in Section~\ref{sec:con}.

\section{Data}\label{sec:data}
We used a combination of astrometric data and low-precision and high-precision RV data for GQ~Lup~B from various sources. In this paper, we present four new epochs of VLTI/GRAVITY astrometry, described in section \ref{GRAVITY}, and also supplement this data with literature astrometry from HST, ALMA, SPHERE, NACO, and MUSE, as mentioned in Table \ref{tab:astrometry}, to extend the time baseline. 

\begin{table*}[ht]
\centering
\caption{Observation Log for GQ Lup}
\begin{tabular}{llcccccccc}
\hline
Object & Date & Start& End & DIT/NDIT/NEXP & airmass & $\tau_0$ & Seeing & Axis & Station \\
 &  & (UT) & (UT) &  &  & (ms) & ('') &  \\
\hline
GQLup\,B & 2021-08-27 & 23:16:16 & 23:25:04 & 30.0\,s / 16 / 2 & 1.1/1.1 &  3.0/6.8 & 0.62/0.73 & OFFAXIS & UTs \\
HD174536\,AB & 2021-08-27 & 23:46:32 & 23:59:51 & 3.0\,s / 12 / 4 & 1.0/1.1 &  4.9/6.6 & 0.64/0.93 & OFFAXIS & UTs \\
GQLup\,AB & 2022-08-14 & 00:48:23 & 01:31:26 & 100.0\,s / 4 / 4 & 1.1/1.2 &  2.5/5.0 & 0.52/0.67 & ONAXIS & ATs \\
GQLup\,AB & 2022-09-06 & 00:11:39 & 00:50:29 & 100.0\,s / 4 / 4 & 1.2/1.4 &  3.8/5.8 & 0.58/1.17 & ONAXIS & ATs \\
GQLup\,AB & 2023-03-19 & 05:48:15 & 06:20:36 & 100.0\,s / 4 / 3 & 1.2/1.3 &  4.8/6.5 & 0.88/1.17 & ONAXIS & ATs \\
\hline
\end{tabular}
\label{tab:obslog}
\end{table*}

\subsection{GRAVITY DATA}\label{GRAVITY}

We observed GQ Lup B with GRAVITY at four different epochs. We used the instrument with combined polarization and medium resolution (R$\approx 400$). The log of the observations are given in Table~\ref{tab:obslog}.
During the night of 2021-08-27, we used GRAVITY on the 8~meter Unit Telescopes (UT), while during the other three nights, we used the 1.8~m Auxiliary Telescopes (AT). When observing with the UTs, the large separation ($\ge$0.6’’) prevented us from using the ON-AXIS mode. Therefore, we used the OFF-AXIS mode and a binary star to calibrate the metrology. This binary was HD17453 \citep{Nowak2024}, a system with a contrast of 40 and a separation of 1.5''. During the four other observation campaigns, we used the ON-AXIS mode and therefore the star itself (GQ Lup A) for metrology referencing.
\begin{figure*}[ht]
    \centering
    \includegraphics[width=\linewidth]{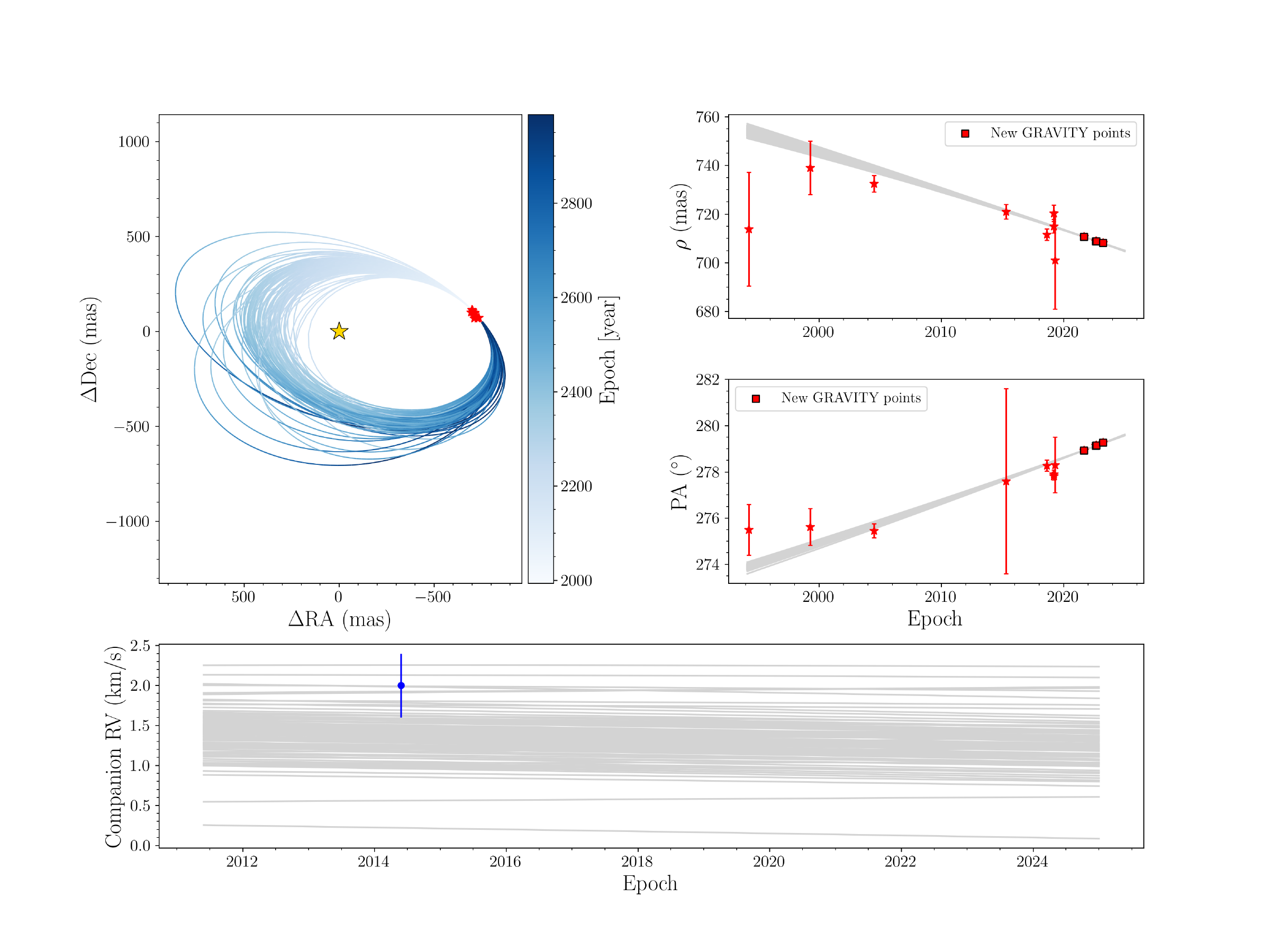}
    \caption{Orbit fit for Case \texttt{GRAVLowRV} (see Table~\ref{tab:fit_cases}), using astrometry including GRAVITY data and low-precision radial velocity measurement. The plot shows posterior samples from the orbit fit for GQ~Lup~B, generated using \texttt{orbitize!}. Blue ellipses represent 100 orbits randomly drawn from the posterior, and red points indicate the relative astrometric data. In the right panels, red points correspond to astrometric measurements from both the literature and new GRAVITY data included in the fit; gray lines trace the same 100 posterior orbits in separation and position angle. The results illustrate the power of GRAVITY data in improving the orbit solution for GQ~Lup~B.}
    \label{fig:orbit}
\end{figure*}
The data were pre-processed by the official ESO GRAVITY pipeline\footnote{Available at \url{https://www.eso.org/sci/software/pipelines/gravity/gravity-pipe-recipes.html}}, yielding un-calibrated  intermediary data products called \texttt{astrored}. 
This product was then fed into a set of GRAVITY consortium Python scripts,  \texttt{run\_gravi\_astrored\_astrometry}\footnote{developed in the context of the ExoGRAVITY large program, and available at \url{https://version-lesia.obspm.fr/repos/DRS_gravity/gravi_tools3/}}, to obtain spectra and astrometry for each epoch. We note that this paper focuses only on the astrometry, while the spectra will be published in a separate paper (Kammerer et al. 2025, in preparation).

A summary of the literature astrometry is presented in Table~\ref{tab:astrometry}, while the new GRAVITY astrometry is provided in Table~\ref{tab:gravity_offsets}.
\begin{table*}[ht]
\centering
\caption{Astrometric Measurements for GQ Lup B}
\begin{tabular}{cccccc}
\hline
MJD & Sep.\ (mas) & Unc.\ (mas) & P.A.\ (deg) & Unc.\ (deg) & Instrument \\
\hline
49445    & 713.8  & 35.5  & 275.5  & 1.1   & ESO 3.6 ComeOn+\textsuperscript{1} \\
51279    & 739  & 11  & 275.62  & 0.86   & HST/WFPC2 \textsuperscript{1} \\
53182    & 732.5  & 3.4  & 275.45  & 0.30   & VLT/NACO \textsuperscript{1} \\
57128    & 721.0  & 3.0  & 277.6  & 4.0   & ALMA\textsuperscript{2} \\
58345.1  & 711.6  & 2.4  & 278.27 & 0.24  & SPHERE\textsuperscript{2} \\
58551.3  & 720.3  & 3.4  & 277.9  & 0.18  & NACO\textsuperscript{3} \\
58557.2  & 715.0  & 2.8  & 277.82 & 0.17  & NACO\textsuperscript{3} \\
58593.1  & 701.0  & 20.0 & 278.3  & 1.2   & MUSE\textsuperscript{3} \\

\hline
\end{tabular}
\vspace{1mm}
\begin{flushleft}
\footnotesize
\textsuperscript{1}\citet{Janson2006};
\textsuperscript{2}\citet{Wu2017}; \textsuperscript{3}\citet{Stolker2021}.
\end{flushleft}
\label{tab:astrometry}
\end{table*}

\begin{table*}[ht]
\centering
\caption{New GRAVITY Astrometric Constraints for GQ Lup B}
\begin{tabular}{cccccc}
\hline
Epoch (MJD) & RA Offset (mas) & RA Unc.\ (mas) & Dec Offset (mas) & Dec Unc.\ (mas) & RA/Dec Corr \\
\hline
$60022.256$  & $-698.969$ & $0.055$ & $114.058$  & $0.153$ & $-0.843$ \\
$59828.023$  & $-699.789$ & $0.060$ & $112.615$  & $0.057$ & $0.652$  \\
$59805.051$  & $-699.876$ & $0.044$ & $112.491$  & $0.062$ & $0.796$  \\
$59453.975$  & $-702.069$ & $0.027$ & $110.363$  & $0.057$ & $0.208$  \\

\hline
\end{tabular}
\label{tab:gravity_offsets}
\end{table*}

\subsection{CRIRES+ DATA}\label{CRIRES}
High-precision RV data were taken from \citet{Gonzalez2025} who report a companion velocity of
\[
v_{\mathrm{B}} = 2.03 \pm 0.04\ \mathrm{km\,s^{-1}},
\]  
measured using the upgraded CRIRES+ instrument on VLT. CRIRES+ offers approximately ten times broader K-band coverage (0.95--5.3$\upmu$m), significantly higher throughput, and resolving power ($R \gtrsim 100{,}000$; \citealp{HolmbergMadhusudhan2022}), as compared to CRIRES.

This value agrees with the earlier, lower-precision RV value from \citep{Schwarz2016}
\[
v_{\mathrm{B}} = 2.0 \pm 0.4\ \mathrm{km\,s^{-1}},
\]  
taken with CRIRES. The larger uncertainty in that result was driven not only by narrower wavelength coverage but also by lower throughput and S/N, yielding ~20–30 in companion spectra only after long exposures. Therefore, the enhanced RV precision reported in \citet{Gonzalez2025} reflects both the broadened spectral coverage of CRIRES+ and its significantly improved throughput and data quality.
However, \citet{Gonzalez2025} acknowledge that their reported uncertainty does not account for additional observed variability in the host star’s RV, estimated by \citet{Donati2012}, to be at the level of  0.4~$\mathrm{km\,s^{-1}}$ between 2009 and 2011. Instrumental effects, intense magnetic stellar activity on the young host star, accretion-driven RV jitter, or an additional unresolved companion in the system may explain this variability. Given this, the RV uncertainty reported in \citet{Gonzalez2025} likely underestimates the true error budget. Additionally, recent work by \citet{Horstman2024} demonstrates the application of this technique to GQ Lup B, using high-resolution KPIC spectroscopy to measure the absolute RVs and search for exomoon-induced modulations. Since the analysis is focused on the planet's RV alone, rather than the relative RV of the planet and the host star, it is more susceptible to instrumental systematics. They report RV variability at the $\sim$400--1000~m\,s$^{-1}$ level, which likely reflects the precision limits of KPIC rather than confirmed RV variability. This highlights the challenges of extracting high-precision RVs from directly imaged companions and the need for caution when interpreting such measurements.
\begin{figure*}[ht]
    \centering
    \includegraphics[width=0.95\textwidth]{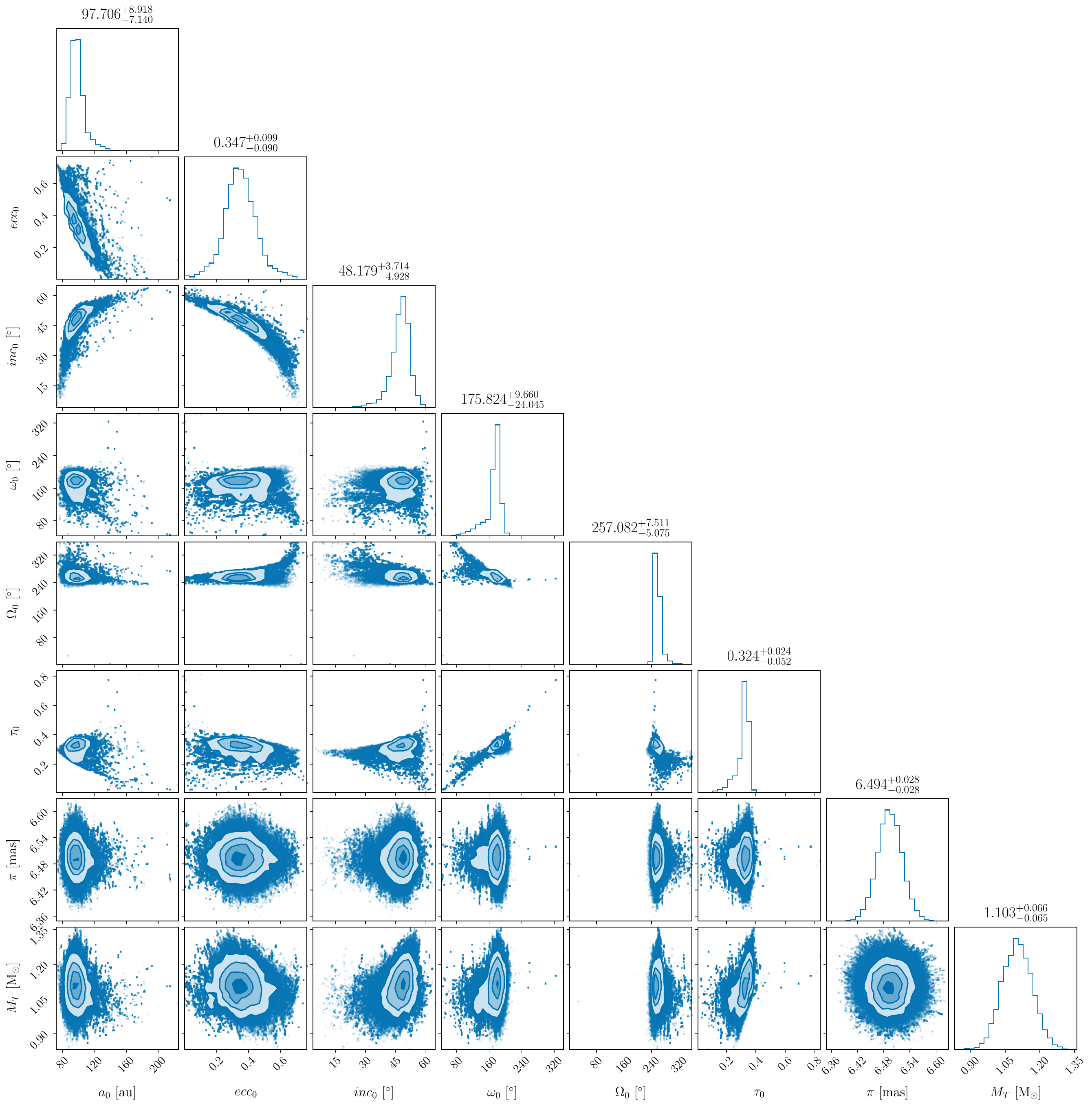} 
    \caption{
    Corner plot of the posterior distributions for Case~{\texttt{GRAVLowRV}} (see Table~\ref{tab:fit_cases}), which includes all available astrometry (including GRAVITY) and low-precision radial velocity measurement. The contours represent the 1-, 2-, and 3-$\sigma$ credible intervals, with individual posterior samples beyond the 3$\sigma$ level shown as blue points. Strong covariances between eccentricity, inclination, and semimajor axis are apparent.
    }
    
    \label{fig:corner}
\end{figure*}
 Given the potential for significant astrophysical variability not reflected in the formal error budget of the high-precision CRIRES+ measurements, we adopt the relative RV value from CRIRES \citep{Schwarz2016} in our analysis, while also exploring how the higher-precision CRIRES+ RV might further constrain the eccentricity of the system. Although the CRIRES measurement is less precise, it benefits from being a differential RV—measuring the planet’s velocity relative to the host star—using spectra taken simultaneously. This approach helps mitigate the impact of instrumental systematics, as shared systematics between the star and planet spectra are largely canceled out. However, if CRIRES introduces variability on timescales shorter than the observation itself (e.g., sub-exposure level), such effects could still influence the measurement. Investigating such potential systematics will be an important direction for future work. In the meantime, we conservatively adopt the CRIRES value for our baseline analysis. The RV measurements are listed in Table~\ref{tab:rv_measurement}.

\begin{table}[ht]
\centering
\caption{Radial Velocity Measurements of GQ Lup B}
\begin{tabular}{ccccc}
\hline
Epoch (MJD) & RV (km\,s$^{-1}$) & Unc.\ (km\,s$^{-1}$) & Instrument \\
\hline
$60003$ & $2.03$ & $\pm0.04$ & CRIRES+\textsuperscript{1} \\
$56806$ & $2.0$ & $\pm0.4$  & CRIRES\textsuperscript{2} \\
\hline
\end{tabular}
\vspace{1mm}
\begin{flushleft}
\footnotesize
\textsuperscript{1}\citet{Gonzalez2025}; \textsuperscript{2}\citet{Schwarz2016}.
\end{flushleft}
\label{tab:rv_measurement}
\end{table}

\section{Orbit Analysis}\label{sec:orbit}
We used {\texttt{\texttt{orbitize}! \citep{Blunt2020}} to perform six distinct orbit fits using various combination of RV and astrometric data. These configurations were designed to evaluate each dataset's individual and combined contributions to the final orbital constraints. The six explored cases are shown in Table \ref{tab:fit_cases}. These combinations were chosen to isolate the effects of GRAVITY astrometry and the precision of the RV data on breaking degeneracies in key orbital elements, such as inclination and eccentricity.

\begin{table*}[ht]
\caption{Overview of Data Set Combinations Used for the Orbital Fitting}
\begin{tabular}{@{}clcccc@{}}
\toprule
Code & Data Used & Lit.\ Astrometry & GRAVITY&  RV  \\
 &  &  & (This Work) &    \\
\midrule
{\texttt{NoGRAVNoRV}} & Astrometry only (no GRAVITY, no RV) & \checkmark & -- & -- \\
{\texttt{GRAVNoRv}} & Astrometry + GRAVITY (no RV)        & \checkmark & \checkmark & --  \\
{\texttt{NoGRAVLowRV}} & Astrometry (no GRAVITY) + Low-precision RV\textsuperscript{a}    & \checkmark & -- & \checkmark \\
{\texttt{NoGRAVHighRV}} & Astrometry (no GRAVITY) + High-precision RV\textsuperscript{b}   & \checkmark & -- & \checkmark  \\
{\texttt{GRAVLowRV}} & Astrometry + GRAVITY + Low-precision RV (\textbf{Adopted})\textsuperscript{a}       & \checkmark & \checkmark & \checkmark \\
\texttt{GRAVHighRV} & Astrometry + GRAVITY + High-precision RV\textsuperscript{b}      & \checkmark & \checkmark & \checkmark  \\
\bottomrule
\end{tabular}
\begin{flushleft}
\footnotesize
\textsuperscript{a} From \citet{Schwarz2016}; \textsuperscript{b} From \citet{Gonzalez2025}.
\end{flushleft}
\label{tab:fit_cases}
\end{table*}

All fits used the following orbital parameters as the model basis: semimajor axis ($a$), eccentricity ($e$), inclination ($i$; where $i = 0^\circ$ corresponds to a face-on orbit), argument of periastron of the companion ($\omega_P$), position angle of nodes ($\Omega$), and the epoch of periastron passage $\tau$, parameterized as a fraction of the orbital period past a reference epoch, defined to be 58849 MJD. We adopted identical uniform priors for all six fits, following \citet{Blunt2020}. We performed all orbital fits using the parallel-tempered implementation of the affine-invariant Markov Chain Monte Carlo (MCMC) sampler \texttt{ptemcee} \citep{ForemanMackey2013, Vousden2016}. For each run, we used 20 temperatures and 1000 walkers. For the fits incorporating RVs or GRAVITY constraints, including the \texttt{GRAVNoRV}},{\texttt{GRAVLowRV}}, and {\texttt{GRAVHighRV}} cases (see Table \ref{tab:fit_cases}), we ran the sampler for 200{,}000 total steps, following an initial burn-in phase of 100,000 steps. For the {\texttt{NoGRAVNoRV}}, \texttt{NoGRAVLowRV}, and {\texttt{NoGRAVHighRV}} fits, we used a shorter chain length of 50{,}000 steps, with 10{,}000 steps for burn-in. In all cases, every 100\textsuperscript{th} step was saved, and the posterior samples were taken from the post-burn-in portion of each walker’s chain. Convergence was assessed by eye.

We imposed Gaussian priors on the total system mass and parallax, with values of M=$1.05\pm0.07$$M_\odot$ \citep{Schwarz2016} and $\pi$=$6.489\pm 0.029$ mas \citep{GAIADR32023}. These parameters do not strongly correlate with any other orbital elements in the posterior, and their marginalized posteriors closely reproduce the imposed priors. 

Figures~\ref{fig:orbit} and~\ref{fig:corner} show one example of the orbit fitting results from case \texttt{GravLowRV}. The orbit plot displays 100 orbital solutions drawn from the posterior, projected onto the sky. The color gradient encodes time, showing the evolution of the planet's apparent motion. Red points on the right indicate the astrometric data points, and the time evolution of the projected separation angle and position angle is shown in gray. The corner plot displays marginalized 1-D and 2-D posterior samples for the eight fitted orbital parameters for case \texttt{GravLowRV}. 

\section{Results}\label{sec:res}
\begin{figure*}[ht]
    \centering
    \includegraphics[width=\textwidth]{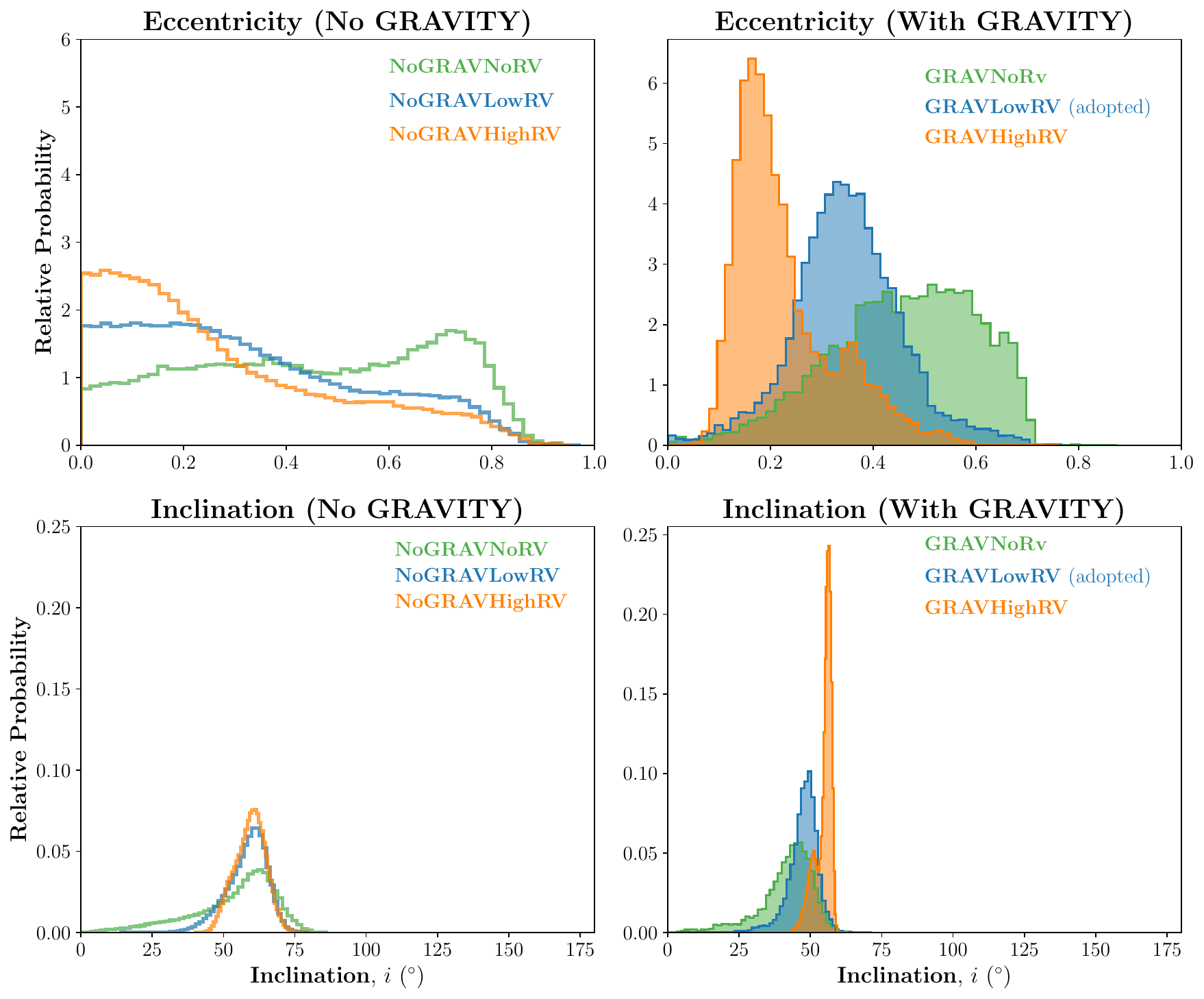} 
    \caption{Posterior eccentricity and inclination distributions for GQ Lup B based on six orbital fits (see Table~\ref{tab:fit_cases}), grouped by inclusion of GRAVITY astrometry. The histograms show the relative probability of eccentricity and inclination values. Inclusion of GRAVITY and relative RV measurements substantially tightens the constraints. While the high-precision RV case (orange) provides the narrowest distribution, it may be biased due to stellar variability or unresolved companions. The adopted blue case ({\texttt{GRAVLowRV}}) incorporates GRAVITY astrometry and low-precision RV, offering a robust and physically plausible solution consistent with a moderately eccentric and inclined orbit.}
    \label{fig:ecc}
\end{figure*}
Our orbit fits reveal that including GRAVITY astrometry substantially improves constraints on the eccentricity and inclination of GQ~Lup~B, as shown in Figure~\ref{fig:ecc}. Without GRAVITY data (cases {\texttt{NoGRAVNoRV}}, {\texttt{NoGRAVLowRV}}, and {{\texttt{NoGRAVHighRV}}}), the eccentricity posteriors are broad and weakly constrained, offering limited insight into the eccentricity of the orbit. In contrast, the GRAVITY-only astrometry case ({\texttt{GRAVNoRV}}) imposes a strong upper limit on the eccentricity, significantly tightening the distribution even to $e = 0.47^{+0.14}_{-0.16}$ even in the absence of RV data, although some inclination-eccentricity degeneracy remains. 

Incorporating radial velocity data further refines the planet’s orbital eccentricity. We adopt the lower-precision measurement from \citet{Schwarz2016}, as the higher-precision dataset may be affected by unmodeled stellar variability or an unresolved companion (see Section~\ref{sec:data}). With the \citet{Schwarz2016} RV included, we obtain $e = 0.35^{+0.10}_{-0.09}$. Although using the high-precision RV data yields a tighter peak at $e=0.20^{+0.14}_{-0.06}$, we treat this value with caution and base our analysis on the more conservative solution from the lower-precision measurement. Similarly, the inclination constrains, visible in the lower panels of Figure \ref{fig:ecc} are broad assuming only literature astrometry, spanning from nearly face-on to significantly inclined orbits. However, with GRAVITY astrometry alone, the inclination peaks at $44^{+6.42}_{-9.62}$$^\circ$. Adding the low-precision RV data refines the value further, resulting in a more sharply peaked inclination posterior at $48^{+3.7}_{-4.9}$$^\circ$. Including the high-precision RV  sharpens the distribution even more, yielding a peak at $56^{+1.5}_{-4.3}$$^\circ$, although we adopt the value derived from the low-precision data for this study as previously stated.

Figure~\ref{fig:contours} presents the joint posterior distributions of eccentricity and inclination for GQ~Lup~B across four modeling configurations: {{\texttt{NoGRAVNoRV}}, {{\texttt{GRAVNoRV}}, {\texttt{GravLowRV}}, and {\texttt{GravHighRV}}. Contours represent the 1$\sigma$ and 2$\sigma$ confidence regions. Including GRAVITY data significantly reduces the volume of the eccentricity–inclination posterior, but a strong degeneracy between the two parameters is still apparent. Including the relative RVs, on the other hand, reduces the degeneracy itself. The {{\texttt{GRAVLowRV}}} configuration, our adopted case, yields a well-localized, unimodal posterior consistent with a moderately eccentric orbit.
 
\begin{figure}[ht]
    \centering
\includegraphics[width=0.46\textwidth]{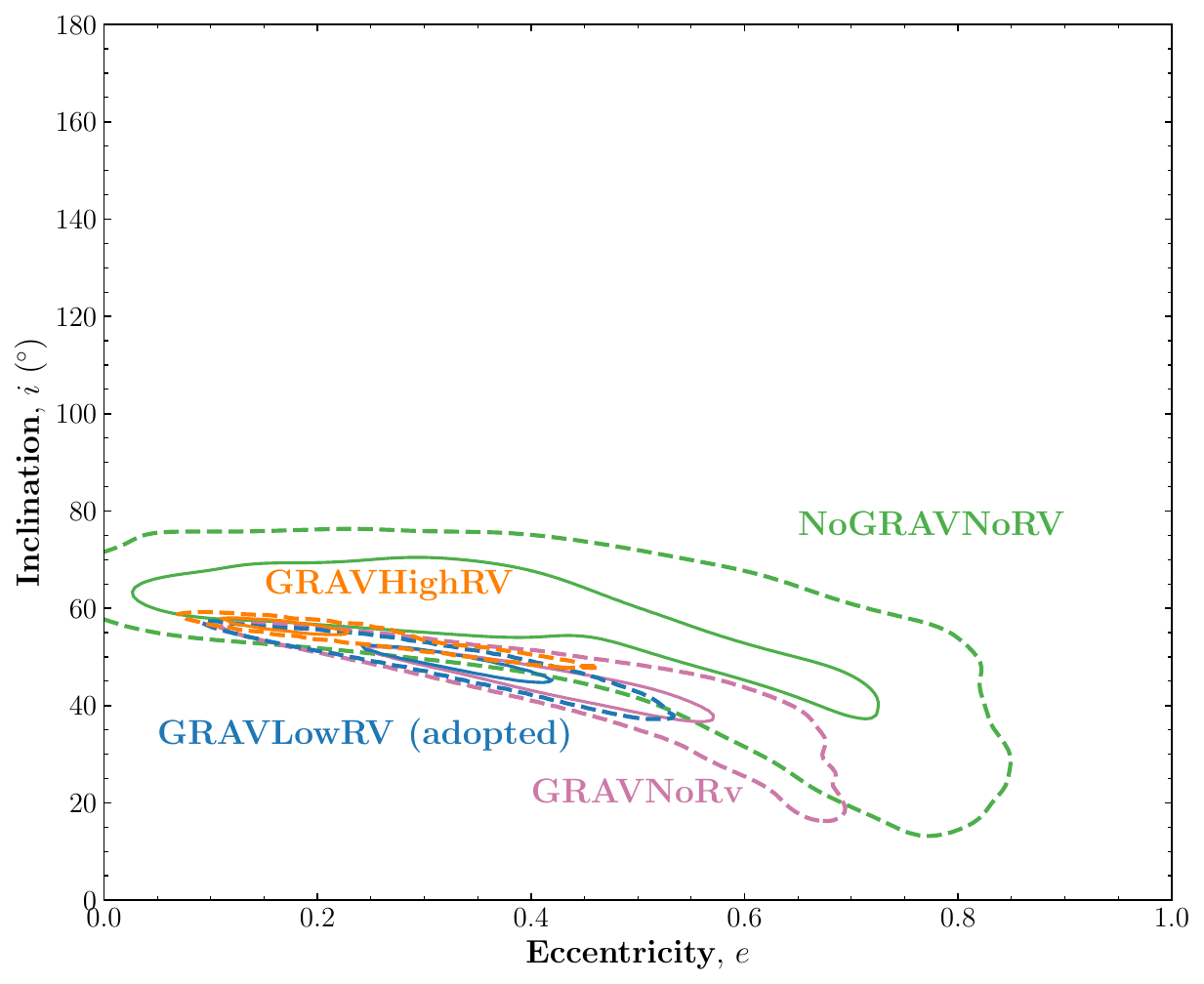}
    \caption{1$\sigma$ and 2$\sigma$ joint posterior distributions of eccentricity and inclination for GQ~Lup~B under four cases (see Table~\ref{tab:fit_cases}) are shown by lines and dashed lines, respectively. The inclusion of GRAVITY astrometry significantly narrows the orbital parameter space, particularly by reducing the volume of the joint eccentricity-inclination posterior. Adding radial velocity measurements—especially in the {{\texttt{GRAVLowRV}}} and {{\texttt{GRAVHighRV}}} cases—further sharpens the constraints by reducing the eccentricity-inclination degeneracy, yielding a well-localized solution consistent with a moderately eccentric orbit.}
    \label{fig:contours}
\end{figure}

Figure~\ref{fig:mu_inc} presents the posterior distribution of the orbital inclination, longitude of ascending node ($\Omega$), and mutual inclination between GQ~Lup~B and the star, the circumstellar disk, and the disk of GQ~Lup~C. In panel 1, the inclination constraints obtained from our adopted case, $\sim48^{+3.7}_{-4.9}$$^\circ$, is compared against the inclination of the circumstellar disk ($\sim60\pm 0.5^\circ$; \citealt{MacGregor2017}), the stellar spin axis ($\sim27\pm 5^\circ$; \citealt{Broeg2007}), and the disk of GQ~Lup~C ($\sim44 \pm 2 ^\circ$; \citealt{Lazzoni2020}). The planet’s orbital inclination appears roughly consistent with that of GQ~Lup~C’s disk.

 In panel 2, the longitude of the ascending node ($\Omega$) is strongly constrained for GQ~Lup~B at $257^{+7.5}_{-5.1}$$^\circ$, and appears offset from both the disk of GQ~Lup~C disk, which peaks at {$315\pm4$ $^\circ$ \citep{Lazzoni2020} and the circumstellar disk which peaks at {$346\pm1$$^\circ$ \citep{MacGregor2017}. The star's $\Omega$ is unknown and is assumed to be uniform between 0 and 360$^\circ$ as shown by the yellow line. 
 
We use equations from \citep{Bowler2023} to calculate the mutual inclination between our adopted planet fit and the circumstellar disk, the stellar spin axis and also the disk of GQ~Lup~C. Panel 3 shows that GQ~Lup~B appears misaligned with both the disk by $63^{+6}_{-14}$$^\circ$ and $52^{+19}_{-24}$$^\circ$ with the stellar spin, confirming a misaligned configuration. The mutual inclination between GQ~Lup~B and the disk of  GQ~Lup~C is smaller, peaking at $35^{+6}_{-13}$$^\circ$.

\begin{figure*}[ht]
    \centering
    \includegraphics[width=0.9\textwidth]{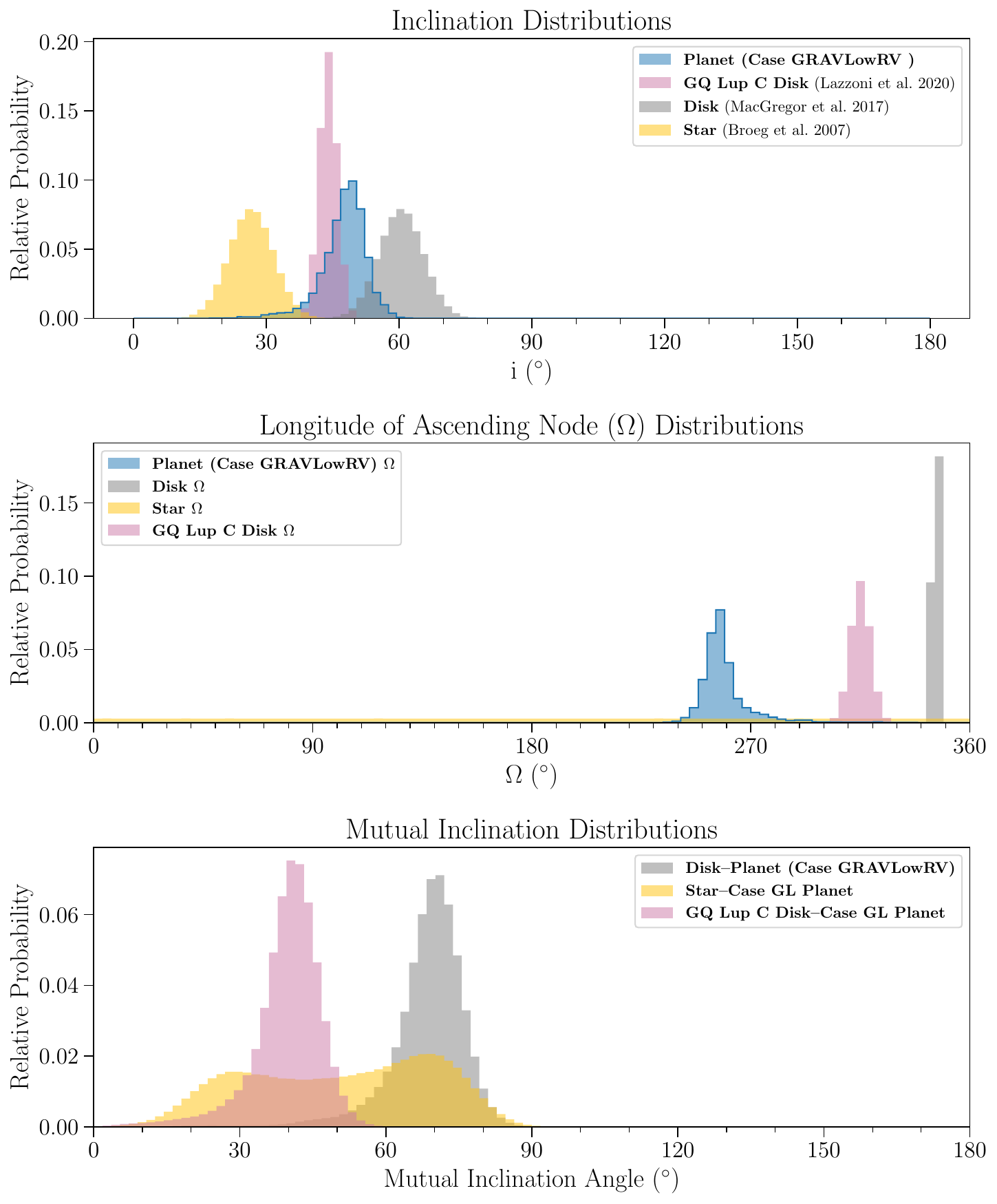} 
    \caption{Inclination and longitude of ascending node ($\Omega$) distributions for the {{\texttt{GRAVLowRV}}}-case orbit of GQ Lup B (blue), compared against the circumstellar disk (\citealt{MacGregor2017}), stellar spin axis (\citealt{Broeg2007}), and GQ Lup C’s disk (\citealt{Lazzoni2020}). The final panel shows the mutual inclination angle between the planet and each reference plane.}
    \label{fig:mu_inc}
\end{figure*}

\begin{table*}[ht]
\centering
\resizebox{\textwidth}{!}{%
\begin{tabular}{@{} lcccccc @{} }
\toprule
\textbf{Parameter} & \textbf{{\texttt{NoGRAVNoRV}}} & \textbf{{\texttt{GRAVNoRV}}} & \textbf{{\texttt{NoGRAVLowRV}}} & \textbf{{\texttt{NoGRAVHighRV}}} & \textbf{{\texttt{GRAVLowRV}}} & \textbf{{\texttt{GRAVHighRV}}} \\
\midrule
SMA [au] & $112.45^{+74.30}_{-34.25}$ & $98.94^{+11.43}_{-10.80}$ & $112.97^{+28.38}_{-22.17}$ & $114.25^{+22.93}_{-17.61}$ & $97.71^{+8.92}_{-7.14}$ & $106.55^{+5.59}_{-5.53}$ \\
$e$ & $0.46^{+0.27}_{-0.29}$ & $0.47^{+0.14}_{-0.16}$ & $0.28^{+0.31}_{-0.19}$ & $0.21^{+0.32}_{-0.15}$ & $0.35^{+0.10}_{-0.09}$ & $0.20^{+0.14}_{-0.06}$ \\
$i$ [deg] & $57.43^{+9.40}_{-19.18}$ & $43.52^{+6.42}_{-9.62}$ & $59.24^{+5.60}_{-8.22}$ & $59.73^{+4.97}_{-6.16}$ & $48.18^{+3.71}_{-4.93}$ & $55.67^{+1.45}_{-4.32}$ \\
$\omega_0$ [deg] & $190.94^{+96.26}_{-134.68}$ & $196.51^{+20.05}_{-23.57}$ & $166.23^{+116.69}_{-69.14}$ & $164.14^{+132.94}_{-57.28}$ & $175.82^{+9.66}_{-24.05}$ & $161.25^{+13.53}_{-37.91}$ \\
$\Omega_0$ [deg] & $199.04^{+69.34}_{-135.95}$ & $241.01^{+15.53}_{-23.66}$ & $266.54^{+21.09}_{-11.36}$ & $265.51^{+20.25}_{-5.91}$ & $257.08^{+7.51}_{-5.08}$ & $263.74^{+8.81}_{-2.25}$ \\
$\tau_0$ & $0.36^{+0.26}_{-0.16}$ & $0.30^{+0.03}_{-0.05}$ & $0.37^{+0.36}_{-0.19}$ & $0.37^{+0.41}_{-0.19}$ & $0.32^{+0.02}_{-0.05}$ & $0.33^{+0.04}_{-0.12}$ \\
$\pi$ [mas] & $6.49^{+0.03}_{-0.03}$ & $6.49^{+0.03}_{-0.03}$ & $6.49^{+0.03}_{-0.03}$ & $6.49^{+0.03}_{-0.03}$ & $6.49^{+0.03}_{-0.03}$ & $6.51^{+0.03}_{-0.03}$ \\
$M_\mathrm{tot}$ [$M_\odot$] & $1.05^{+0.07}_{-0.07}$ & $1.06^{+0.07}_{-0.07}$ & $1.05^{+0.07}_{-0.07}$ & $1.04^{+0.07}_{-0.07}$ & $1.10^{+0.07}_{-0.06}$ & $1.28^{+0.07}_{-0.15}$ \\
\bottomrule
\end{tabular}%
}
\caption{Posterior medians and $1\sigma$ uncertainties for orbital parameters under six fitting configurations. Parameters shown are: 
$a$—semi-major axis [au]; $e$—eccentricity; $i$—inclination [deg]; $\omega_P$—argument of periastron [deg]; 
$\Omega$—longitude of ascending node [deg]; $\tau_0$—normalized epoch of periastron; 
$\pi$—parallax [mas]; and $M_\mathrm{tot}$—total system mass [$M_\odot$] for all 6 cases.}
\label{tab:orbital_fits}
\end{table*}

\section{Discussion}\label{sec:dis} 
Our results are consistent with the eccentricity and inclination calculations from \citet{Stolker2021} where they combine RV measurements from \citep{Schwarz2016} with astrometry from VLT/NACO and VLT/MUSE. They report an eccentricity and inclination constraint of $0.24^{+0.32}_{-0.17}$ and $60^{+5}_{-9}$ compared to $0.35^{+0.10}_{-0.09}$ and $48.2^{+3.7}_{-4.9}$ in this work. Moreover, they find a mutual inclination of $84^{+9}_{-9}$ between the orbit of the companion and the circumstellar disk compared to our constraint of $63^{+6}_{-14}$. By incorporating high-precision GRAVITY astrometry, we have tightened these constraints with reduced errors on both eccentricity and inclination compared to \citet{Stolker2021}.

The formation pathway for GQ~Lup~B has remained elusive since its discovery. \cite{Neuh2005} argued for a formation scenario consistent with cloud formation, which appears consistent with the presence of a compact circumstellar disk imaged by ALMA; such a disk could be too small to have formed GQ~Lup~B in-situ  \citep{MacGregor2017}. More recently, \citep{Stolker2021} found the disk to be significantly misaligned with the orbit, suggesting a history of dynamic interactions or cloud fragmentation. Complementary to the architectural constraints, the consistent chemical compositions of GQ Lup A and B point towards formation via cloud fragmentation \citep{Stamatellos2007, Stam&Whitworth2009}. Our analysis leverages the power of VLT/GRAVITY astrometry and high-precision spectroscopy from CRIRES+ to provide precise and robust estimates of the orbit and favors the expectations from dynamically hot in-situ formation processes like cloud fragmentation.

 We have shown that when we combine high-precision astrometry from GRAVITY and high-precision spectroscopy from CRIRES+, we can precisely constrain 3-D orbits of planets at $\sim$100~au. For the case of GQ~Lup~B, the addition of planetary RVs—despite our cautious adoption of the lower-precision dataset—has significantly improved constraints on the planet’s orbit. The moderate eccentricity ($e\sim0.35$) and well-localized inclination (near $48^{+3.7}_{-4.9}$ $^\circ$) are now less strongly degenerate. The resulting orbital solution places the planet between the inner circumstellar disk (truncated at $\sim$50–75~au) and the wide companion GQ~Lup~C (projected separation $\sim$2400~au).

The orbital architecture of the GQ~Lup system points to a dynamically complex history. The orbital plane of GQ~Lup~B (i=$\simeq48.6^\circ$) is likely misaligned with both the stellar spin axis (i$\simeq26.6^\circ$) and the circumstellar disk (i$\simeq60.5^\circ$), disfavoring a formation via coplanar, in-situ core accretion. Interestingly, although GQ~Lup~B is misaligned with the primary disk, its inclination is roughly consistent with that of GQ~Lup~C. The similarity in inclination and longitude of ascending node---though still uncertain---experienced similar early dynamical shaping. Dynamical simulations of the systems formation could help shed light on which of these scenarios is most likely

Using the posterior distributions from our adopted orbital solution (Fit {{\texttt{GRAVLowRV}}), we find a pericenter distance of $65^{+15}_{-14}$~au. The disk, as observed by ALMA and presented in \citet{MacGregor2017}, appears smooth and symmetric out to an outer radius of $\sim$50~au. Additionally, the polarimetry results also show potential spiral structure in the disk \citep{VanHolstein2021}, which {\color{blue} can be caused due to} the pericenter distance. We estimated the half-width of the chaotic zone can be estimated using the expression from \citet{Morrison2015},
\begin{equation}
    \Delta a \approx 1.5 \, \mu^{0.28} \, a_p,
\end{equation}
where
\begin{itemize}
    \item $\Delta a$ is the half-width of the chaotic zone,
    \item $a_p$ is the semi-major axis of the companion,
    \item $\mu = \frac{M_p}{M_*}$ is the mass ratio of the companion ($M_p$) to the host star ($M_*$).
\end{itemize}
We find the chaotic zone to be extend from 49.8 AU to 150.2 AU, indicating that the companion can dynamically truncate or sculpt the disk. Our calculated pericenter distance of $65^{+15}_{-14}$~au lies well within this zone, and because the polarimetry reveals a warp at similar radii, it is possible that the companion is truncating and warping the inner edge of the disk.

We also assessed the plausibility of Kozai--Lidov (KL) oscillations induced by the outer companion GQ~Lup~C. Assuming a circular orbit at 2300~au and component masses of 1\,M$_\odot$ (GQ~Lup~A), 20\,M$_\mathrm{Jup}$ (GQ~Lup~B), and 0.15\,M$_\odot$ (GQ~Lup~C), we estimate a KL timescale of approximately 82~Myr. This was calculated using the quadrupole-order approximation for hierarchical triples \citep{Antognini2015},

\begin{equation}
    t_{\mathrm{KL}} \approx \frac{M_{\mathrm{A}} + M_{\mathrm{B}}}{M_{\mathrm{C}}} \cdot \frac{P_{\mathrm{C}}^2}{P_{\mathrm{B}}} \cdot (1 - e_{\mathrm{C}}^2)^{3/2},
\end{equation}

where $P_{\mathrm{B}}$ and $P_{\mathrm{C}}$ are the orbital periods of the inner and outer binaries, respectively. This timescale is much longer than the estimated age of the system ($\sim$1-5~Myr), suggesting that secular interactions from GQ~Lup~C are unlikely to have played a significant role shaping its current orbit but could become dynamically relevant over the system’s lifetime.

Dynamical scattering is unlikely. GQ~Lup~B could be the inner survivor of a past interaction that ejected a lower-mass object, possibly on the order of a few Jupiter masses. Conversely, GQ~Lup~B could be the outer object resulting from scattering by an unseen, more massive inner perturber. However, the smooth morphology of the disk argues against the presence of such a companion in the inner few tens of AU. We also find no evidence of a transition disk or inner cavity in the current literature, though further imaging could help clarify this point. If GQ~Lup~A were a close binary, that would likely be visible in the RVs of the primary. Atmospheric constraints also point to dynamical scattering as a less probably scenario.

Flyby interactions are unlikely in this case. While GQ~Lup shares proper motion characteristics with core members of the Lupus star-forming region—suggesting it is not an interloper—this alone does not rule out the possibility of a past stellar encounter. However, the continued presence of the wide tertiary companion, GQ~Lup~C, at a projected separation of $\sim$2400~au places strong constraints on the severity of any potential flyby. A sufficiently close or energetic encounter capable of exciting the observed eccentricity or inclination would likely have disrupted the outer companion, making such a scenario implausible.

The constraints suggest that GQ~Lup~B likely formed via cloud fragmentation with very minimal dynamical reshaping. The KL timescale exceeds the age of the system, making significant excitation unlikely. The inner companion's orbit is misaligned with the disk, consistent with primordial tilt imparted at formation rather than due to later pertubations. The inner companion has a moderate, not extreme, eccentricity, and matched expectation for cloud collapse and argues against violent scattering events. Overall, our calculations support a formation scenario that is dominated by cloud fragmentation.

\section{Conclusions}\label{sec:con} 
In this work, we combine the high-precision astrometric measurements from GRAVITY, SPHERE, and NACO with high-resolution ($R \sim 100{,}000$) spectroscopic data from CRIRES+ to refine the orbit of GQ~Lup~B. While we adopt the lower-precision RV dataset from \citet{Schwarz2016} due to concerns about unmodeled systematics in higher-precision measurements (e.g., stellar activity or an unresolved companion), even this modest precision measurement (2.03 $\pm$ 0.4~$\mathrm{km\,s^{-1}}$)---when combined with interferometric astrometry---proves powerful. It helps break the known degeneracy between inclination and eccentricity in long-period, astrometry-only orbits, allowing us to obtain the first well-constrained eccentricity posterior for GQ~Lup~B. Specifically, the eccentricity is refined from $0.47^{+0.14}_{-0.16}$ (GRAVITY-only) to $0.35^{+0.10}_{-0.09}$ with the inclusion of RVs and GRAVITY data.

Beyond orbital parameter estimation, we explore the 3-D architecture of the system by computing the mutual inclinations between GQ~Lup~B, the stellar spin axis, the circumstellar disk, and the wide companion GQ~Lup~C. We find that GQ~Lup~B is misaligned with the inner system—both the disk and stellar spin axis—but comparatively more aligned with the disk of GQ~Lup~C. This configuration suggests that GQ~Lup~B may have formed via cloud fragmentation, and its misalignment with the outer disk is coincidental.

Our results underscore how the combination of high-precision astrometry and planetary radial velocities enables significantly more robust orbital constraints for wide-orbit companions. This technique is particularly valuable in systems like GQ~Lup, where traditional core accretion models are challenged and observational baselines are short. As high-resolution spectroscopic techniques continue to mature and as instruments like GRAVITY and CRIRES+ expand their reach, we are entering an era where joint astrometry-RV analysis will allow us to uncover population-level trends between inclination and eccentricity, and formation pathways, ultimately building a holistic understanding of the orbits of companions at wide separations.

\begin{acknowledgments}
Based on observations collected at the European Southern Observatory under ESO programmes: PI: Sylvestre Lacour, 1104.C-0651, 109.238N.001 and 109.238N.002. 

V.V.\ acknowledges support from the NASA FINESST Fellowship under award number 80NSSC21K1852, the University of California, Irvine's Graduate Dean's Dissertation Fellowship. V.V. and A.S. gratefully acknowledge partial support from UC Irvine for this work. V.V.\ extends sincere thanks to Ruth Murray-Clay for her valuable insights. V.V.\ would like to thank Paul Robertson and his group members for valuable discussions, as well as the members of Jason Wang's BOBA group. S.L.\ acknowledges the support of the French Agence Nationale de la Recherche (ANR-21-CE31-0017, ExoVLTI) and of the European Research Council (ERC Advanced Grant No. 101142746, PLANETES). J.J.W.\ and A.C.\ are supported by NASA XRP Grant 80NSSC23K0280. Based on observations collected at the European Southern Observatory under ESO programmes 1104.C-0651 and 109.238N.001.

The authors acknowledge the use of OpenAI's ChatGPT for assistance with proofreading and visualization suggestions during the preparation of this manuscript. All scientific content and interpretations come from the authors.
\end{acknowledgments}

\section*{Data Availability}
{The code for generating the plots and additional corner and orbit plots for other 5 cases are available on Zenodo (\url{https://zenodo.org/records/17087542})
and GitHub (\url{https://github.com/astrovidee/GQ_Lup_B_analysis_2025}).}

\facilities{VLTI/GRAVITY \citep{Lacour2020ExoGRAVITY}}

\software{\texttt{numpy} \citep{harris2020array}, \texttt{pandas} \citep{pandas}, \texttt{matplotlib} \citep{Hunter:2007aa}, \texttt{corner} \citep{corner}, \texttt{scipy} \citep{scipy}, \texttt{astropy} (\citealt{astropy:2013}, \citealt{astropy:2018}, \citealt{astropy:2022}), \texttt{\texttt{orbitize}!} \citep{Blunt2020}, \texttt{Aspro}\footnote{Available at \url{http://www.jmmc.fr/aspro}} \citep{2016SPIE.9907E..11B}.}

\textit{For Proofreading}: ChatGPT\citep{openai2023}
          
\bibliography{main}{}
\bibliographystyle{yahapj}
\end{document}